\documentclass{elsart}
\usepackage{graphicx} 
\usepackage{dcolumn}  
\usepackage{bm}       
\usepackage{color}
\usepackage{latexsym}
\usepackage{amsfonts}
\usepackage{dsfont}
\journal{Nuclear Physics A}
\usepackage{latexsym}
\begin{document}
\newcommand{\leftg}{\langle \phi_0 |} 
\newcommand{\rightg}{| \phi_0 \rangle} 
\newcommand{\chiral}{\langle \bar{q} q \rangle} 
\newcommand{\vs}{\vspace{-0.25cm}}
\newcommand{\gev}{\,\mathrm{GeV}}
\newcommand{\mev}{\,\mathrm{MeV}}
\newcommand{\fmd}{\,\mathrm{fm}^{-3}}
\newcommand{\iso}{\vec{\tau}}
\newcommand{\ord}[1]{\mathcal{O}(k_f^{#1})}
\newcommand{\dif}{\mathrm{d}}
\newcommand{\rbot}{r_{\scriptscriptstyle{\bot}}}
\newcommand{\halfx}{{\scriptstyle {1\over 2}}}
\newcommand{\bff}[1]{{\mbox{\boldmath $#1$}}}
\newcommand{\D}{\displaystyle}
\newcommand{\lp}{\left}
\newcommand{\rp}{\right}
\newcommand{\dist}{(\bm{r}_1 - \bm{r}_2)}
\newcommand{\fud}[2]{\frac{\delta #1}{\delta #2}}
\newcommand{\pad}[2]{\frac{\partial #1}{\partial #2}}
\newcommand{\bra}[1]{\left\langle #1 \right|}
\newcommand{\ket}[1]{\left| #1 \right\rangle}
\newcommand{\vev}[2]{\bra{#1} #2 \ket{#1}}
\newcommand{\boma}[1]{\mbox{\boldmath $#1$}}
\newcommand{\identity}{{\sf 1 \hspace{-0.15em}
                        \rule{0.087em}{1.5ex}
                        \rule{0.12em}{0.1ex}
                        \hspace{-0.3em}
                        \rule[1.4ex]{0.12em}{0.1ex}
                        \hspace{0.3em}}}


\begin{frontmatter}
\title{Chiral pion-nucleon dynamics in finite nuclei: 
spin-isospin excitations}
\author[Bo,Mun]{P. Finelli},
\author[Mun]{N. Kaiser},
\author[Vre]{D. Vretenar},
\author[Mun]{W. Weise}
\address[Bo]{Physics Department, University of Bologna, Italy}
\address[Mun]{Physik-Department, Technische Universit\"at M\"unchen, Germany}
\address[Vre]{Physics Department, Faculty of Science, University of
Zagreb, Croatia}
\begin{abstract}
The nuclear density functional framework, based on
chiral dynamics and the symmetry breaking pattern of low-energy QCD, 
is extended to the description of collective nuclear
excitations. Starting from the relativistic point-coupling 
Lagrangian introduced in Ref.~\cite{Fi.06}, the proton-neutron
(quasiparticle) random phase approximation is
formulated and applied to investigate 
the role of chiral pion-nucleon dynamics in excitation modes
involving spin and isospin degrees of freedom, e.g. isobaric analog
states and Gamow-Teller resonances.

\end{abstract}  
\date{\today}
\end{frontmatter}


\section{Introduction.}

In a series of recent articles \cite{Fi.06,Fi.02,Fi.03,VW.04,KNV.05,nucmat3} we
have developed a novel approach to nuclear density functional theory
which emphasizes the relationship between aspects of low-energy, 
non-perturbative 
QCD and the nuclear many-body dynamics. A relativistic nuclear energy
density functional has been formulated, starting from two basic features 
which establish the link with low-energy QCD and its symmetry-breaking 
pattern: (a) strong scalar and vector fields related to in-medium changes 
of QCD vacuum condensates; 
(b) long- and intermediate-range interactions generated by one-and 
two-pion exchange, derived from in-medium chiral perturbation theory, 
with explicit inclusion of $\Delta(1232)$ excitations and in 
combination with three-nucleon (3N) interactions.

The resulting nuclear matter energy density functional~\cite{nucmat3} 
is mapped~\cite{Fi.03} onto the exchange-correlation energy density 
functional of a covariant point-coupling 
model for finite nuclei, including second order gradient corrections. 
In this approach the nuclear ground state is determined by the
self-consistent solution of the relativistic generalizations 
of the linear single-nucleon Kohn-Sham equations. 
The construction of the density functional involves an expansion of 
nucleon self-energies in powers of the Fermi momentum up to and including 
terms of order $k_f^6$. In comparison with purely 
phenomenological mean-field models,
the built-in QCD constraints and the explicit treatment
of pion exchange restrict the freedom in adjusting
parameters and functional forms of the density-dependent couplings. 
This approach to nuclear dynamics has been tested in an
analysis of ground-state properties of a broad range of spherical 
and deformed nuclei \cite{Fi.06}. 
The results have been compared with experimental data on 
binding energies, charge radii, neutron radii and deformation 
parameters for several isotopic chains. It has been demonstrated that the 
results for nuclear ground-state properties are at the same level 
of quantitative comparison with data as the best phenomenological 
relativistic mean-field models.

In this work the low-energy QCD-constrained nuclear density functional 
framework will be extended to the description of excited states and,
in particular, to spin-isospin collective excitations. Starting from the 
relativistic point-coupling Hartree-Bogoliubov model of Ref.~\cite{Fi.06},
we formulate the proton-neutron (quasiparticle) random phase 
approximation (PN-QRPA). This is then applied 
in an analysis of isobaric analog states and
Gamow-Teller resonances in doubly-closed and open-shell nuclei. 
Our aim is to study the role of chiral pion-nucleon dynamics 
in charge-exchange modes. The results will be compared with available 
data, and with QRPA calculations performed using one of the 
modern phenomenological relativistic density-dependent effective 
interactions based on the meson-exchange representation of 
effective nuclear forces.

\section{The Quasiparticle Random Phase Approximation}

The matrix equations of the 
relativistic random phase approximation (RRPA) for an 
effective Lagrangian with explicit density dependence of the 
meson-nucleon couplings have been derived in Ref.~\cite{Niksic:2002ri}, 
and for nucleon-nucleon point-coupling effective Lagrangians 
in Ref.~\cite{Niksic:2005kb}.
The relativistic QRPA for collective excitations in open-shell nuclei
has been formulated in Ref.~\cite{Paar:2002gz} using the canonical   
single-nucleon basis of the relativistic Hartree-Bogoliubov 
(RHB) model. The extension to spin and isospin excitations 
(PN-QRPA) has been described in Ref.~\cite{Paar:2004re}. Our implementation of 
the PN-QRPA is identical to that of Ref.~\cite{Paar:2004re}, 
with the important difference that it is now
based on the point-coupling effective 
Lagrangian with density-dependent couplings determined by 
chiral pion-nucleon dynamics~\cite{Fi.06}, rather than on 
phenomenological meson-exchange effective interactions. Here
we only collect the essential expressions and refer the 
reader to Refs.~\cite{Niksic:2002ri,Niksic:2005kb,Paar:2002gz,Paar:2004re}
for technical details of the relativistic QRPA.
 
We consider transitions between the $0^+$ ground state of a spherical
even-even parent nucleus and a state with angular momentum and parity
$J^\pi$ of the odd-odd daughter nucleus.
The spin-isospin dependent 
coupling terms are generated by the following 
interaction parts of the Lagrangian density
\footnote{Vectors in isospin space are denoted by arrows, and boldface 
symbols indicate vectors in ordinary three-dimensional space.}:
\begin{eqnarray}
\mathcal{L}_{int} & = & \mathcal{L}_{IV} +  
 \mathcal{L}_{\pi N} + \mathcal{L}_{LM} \label{L_int}\\
\mathcal{L}_{IV}  & = &
      - \frac{1}{2}G_{TS}^{(\pi)}(\hat{\rho}) (\bar{\psi} \iso \psi)
      \cdot (\bar{\psi} \iso \psi)
      - \frac{1}{2}G_{TV}^{(\pi)}(\hat{\rho}) (\bar{\psi} \iso \gamma_\mu \psi)
      \cdot (\bar{\psi} \iso \gamma^\mu \psi)  \label{FKVW-IV}\\
\mathcal{L}_{\pi N} & = & - \frac{g_A}{2f_{\pi}} \bar{\psi}\gamma^{\mu}\gamma_{5}
      \vec{\tau} \psi \cdot \partial_{\mu} \vec{\phi}  
      \label{one-pi} \\ 
\mathcal{L}_{LM}  & = &  \frac{1}{2}{g_0'}(\hat{\rho})  
      (\bar{\psi} \iso \gamma_5 \gamma_\mu \psi) \cdot 
      (\bar{\psi} \iso \gamma_5\gamma^\mu \psi) \label{LM} \; .
\label{lagrangian}	
\end{eqnarray}
The first term $\mathcal{L}_{IV}$ collects the isovector parts 
of the nuclear density functional of Ref.~\cite{Fi.06}, with the 
density-dependent coupling strengths $G_{TS}^{(\pi)}$ and
$G_{TV}^{(\pi)}$ completely determined by chiral pion-nucleon dynamics. 
Details are given in Ref.~\cite{Fi.06}. The same approach, 
without further adjustment, will also be used for the proton-neutron 
particle-hole ($ph$) residual interaction. Because of parity conservation, 
the one-pion direct contribution vanishes at the mean-field level in 
the calculation of a nuclear ground state. The pion-nucleon 
Lagrangian density $\mathcal{L}_{\pi N}$ Eq.~(\ref{one-pi}) must  
be included, however, in calculations of excitations that involve
spin and isospin degrees of freedom \cite{wolfram}. 
The axial vector coupling constant: $g_A = 1.3$, and 
the pion decay constant: $f_\pi = 92.4$ MeV.

Short-distance spin-isospin dynamics is encoded in the zero range
Landau-Migdal term $\mathcal{L}_{LM}$, Eq.~(\ref{LM}).
In the non-relativistic limit the corresponding two-body interaction
reduces to the familiar form $g_0'\bm{\sigma}_1 \cdot \bm{\sigma}_2 
\vec{\tau}_1 \cdot \vec{\tau}_2$. In a recent detailed 
study of the role of $2\pi$-exchange in the interaction of 
quasi-nucleons at the Fermi surface 
$|\vec{p}_{1}| = |\vec{p}_{2}| = k_f$ 
\cite{Kaiser:2006wt}, the four Landau-Migdal parameters 
${f_0(k_f)}$, ${f_0'(k_f)}$, ${g_0(k_f)}$ and ${g_0'(k_f)}$ 
which characterize the four spin-isospin channels of the isotropic 
($l=0$) part of the N-N quasiparticle interaction, have been calculated 
using in-medium chiral perturbation theory (ChPT). 
This calculation includes contributions 
from $1\pi$-exchange, iterated $1\pi$-exchange 
and $2\pi$-exchange with virtual $\Delta(1232)$-isobar excitations. 
The resulting dependence on the Fermi momentum $k_f$ is converted into 
density dependence using the relation for symmetric nuclear 
matter: $\rho = 2 k_f^3/(3\pi^2)$. 

In general, the explicit density dependence of the couplings 
introduces additional rearrangement terms in the residual 
two-body interaction of the RRPA \cite{Niksic:2002ri}.
However, since these terms include the corresponding 
isoscalar ground-state densities, it is easy to see 
that they are absent in the charge-exchange channels, 
and the residual two-body interaction reads:
\begin{eqnarray}
\label{Vph}
V^{ph}(\bm{r}_1,\bm{r}_2) & = &
G_{TS}^{(\pi)} (\rho) \; \iso_1 \cdot \iso_2 ~ (\mathds{1})_1  
(\mathds{1})_2 \; \delta^3 \dist \nonumber \\
& ~ & + \; G_{TV}^{(\pi)} (\rho) \; \iso_1 \cdot \iso_2 ~(\beta
\gamma^\mu)_1 (\beta \gamma_\mu)_2 \; \delta^3 \dist \nonumber \\
& ~ & + \; \frac{g_A^2}{16\pi f_\pi^2} ~\iso_1 \cdot \iso_2 
~(\bm{\Sigma}_1 \cdot \bm{\nabla}) (\bm{\Sigma}_2 \cdot
\bm{\nabla}) \frac{e^{-m_\pi |\bm{r}_1 - \bm{r}_2|}}{|\bm{r}_1 -
  \bm{r}_2|} \nonumber \\
& ~ & + \; g_0'(\rho) \; \iso_1 \cdot \iso_2 ~(\beta \bm{\Sigma})_1 \cdot 
(\beta \bm{\Sigma})_2
\; \delta^3 \dist \;.
\end{eqnarray}
Here $\bm \Sigma$ is the $4\times 4$ matrix 
representation of the spin operator.
We note that there is no double counting of the one-pion 
contribution, because the Hartree diagram 
does not contribute to the isotropic part of the quasi-nucleon 
interaction in the long wavelength 
limit, i.e. to $g_0'(\rho)$ \cite{Kaiser:2006wt}.

Pairing interactions in the particle-particle channel
have also been derived using in-medium chiral dynamics~\cite{ChiralPair}.
However, for the convenience of quantitative comparison with previous 
relativistic PN-QRPA calculations for open-shell nuclei 
of Ref.~\cite{Paar:2004re}, and consistent with the model 
that we have developed for ground-state properties in 
Refs.~\cite{Fi.06,Fi.02,Fi.03}, in the $T=1$ channel the
pairing part of the Gogny force is used
\begin{equation}
V^{pp}_{T=1}(\bm{r}_1,\bm{r}_2)=\sum_{i=1,2}e^{-((\mathbf{r}_{1}-
\mathbf{r}_{2})/{\mu _{i}}%
)^{2}} \,(W_{i}~+~B_{i}P^{\sigma }-H_{i}P^{\tau }-M_{i}P^{\sigma }P^{\tau }),
\label{GognyT1}
\end{equation}
with the set D1S \cite{Ber.91} for the parameters $\mu _{i}$, $W_{i}$, $%
B_{i} $, $H_{i}$ and $M_{i}$ $(i=1,2)$, and a short-range repulsive Gaussian 
combined with a weaker longer-range attractive Gaussian in the $T=0$ 
channel:
\begin{equation}
V^{pp}_{T=0}(\bm{r}_1,\bm{r}_2) = 
- V_0 \sum_{j=1}^2 g_j \; {\rm e}^{-((\mathbf{r}_{1}-
\mathbf{r}_{2})/{\mu _{j}}%
)^{2}} \; \hat\Pi_{S=1,T=0}
\quad ,
\label{GaussT0}
\end{equation}
where $\hat\Pi_{S=1,T=0}$ projects onto states with $S=1$ and $T=0$.  
The ranges of the two Gaussians $\mu_1$=1.2\,fm and $\mu_2$=0.7\,fm  
are the same as for the Gogny interaction Eq. (\ref{GognyT1}), 
and the choice of the relative strengths $g_1 =1$ and
$g_2 = -2$ makes the force repulsive at small distances. 
The overall strength parameter of the $T=0$ 
pairing interaction $V_0 = 250$ MeV
is the same as in the relativistic PN-QRPA calculation of 
Gamow-Teller resonances of Ref.~\cite{Paar:2004re}. 
The $T=1$ Gogny interaction Eq.~(\ref{GognyT1}) is also used 
in the pairing channel of the RHB equations which determine the 
ground state of the initial nucleus.
The phenomenological particle-particle interaction 
used here differs only slightly from 
the one derived from in-medium ChPT~\cite{ChiralPair}.
A complete investigation of in-medium pionic fluctuations (in the 
particle-hole and in the particle-particle channel) will be presented
in a forthcoming paper.

The two-quasiparticle configuration space includes states with
both nucleons in discrete bound levels, states with one nucleon in a
bound level and the second nucleon in the continuum, and also states with both
nucleons in the continuum. In addition to configurations built from
two-quasiparticle states of positive energy, the RQRPA configuration space
includes pair-configurations formed from the fully or partially occupied
states of positive energy and the empty negative-energy states from the
Dirac sea. In Refs.~\cite{Paar:2002gz,Paar:2004re} it has been shown  
that the inclusion of configurations built from occupied
positive-energy states and empty negative-energy states is essential 
for the consistency of the relativistic (proton-neutron) QRPA
(current conservation, decoupling of spurious states, sum rules).

The total strength for the transition between the ground state of an 
even-even spherical (N,Z) nucleus and the excited state 
$|\lambda J \rangle$ of the 
corresponding odd-odd (N+1,Z-1) or (N-1,Z+1) 
nucleus, induced by an isovector single-particle operator $T^{JM}$, reads
\begin{equation}
B_{\lambda J}^{\pm} = \lp| \sum_{pn} \langle p||T^J||n \rangle 
\lp( X_{pn}^{\lambda J} u_p v_n + (-1)^J Y_{pn}^{\lambda J}v_p u_n \rp) \rp|^2
\; , 
\label{strength-}
\end{equation}
where $p$ and $n$ denote proton and neutron quasiparticle
canonical states, respectively, with occupation factors 
$u_p$, $v_p$, $u_n$, $v_n$. $X^{\lambda J}$ and $Y^{\lambda J}$ 
are the forward- and backward-going QRPA amplitudes for the 
state $|\lambda J \rangle$. 
The discrete strength distribution is folded with the Lorentzian function
\begin{equation}
R(E)^{\pm} = \frac{1}{\pi}\sum_{\lambda}B_{\lambda J}^{\pm}
\frac{\Gamma/2}{(E-E_{\lambda_{\pm}})^2+(\Gamma /2)^2} \; .
\label{lorentzian}
\end{equation}
In the calculations performed in this work the choice for
the width of the Lorentzian is 1 MeV.

\subsection{Determination of $g_0'(k_f)$.}

\begin{figure}
\includegraphics[scale=0.65,angle=0]{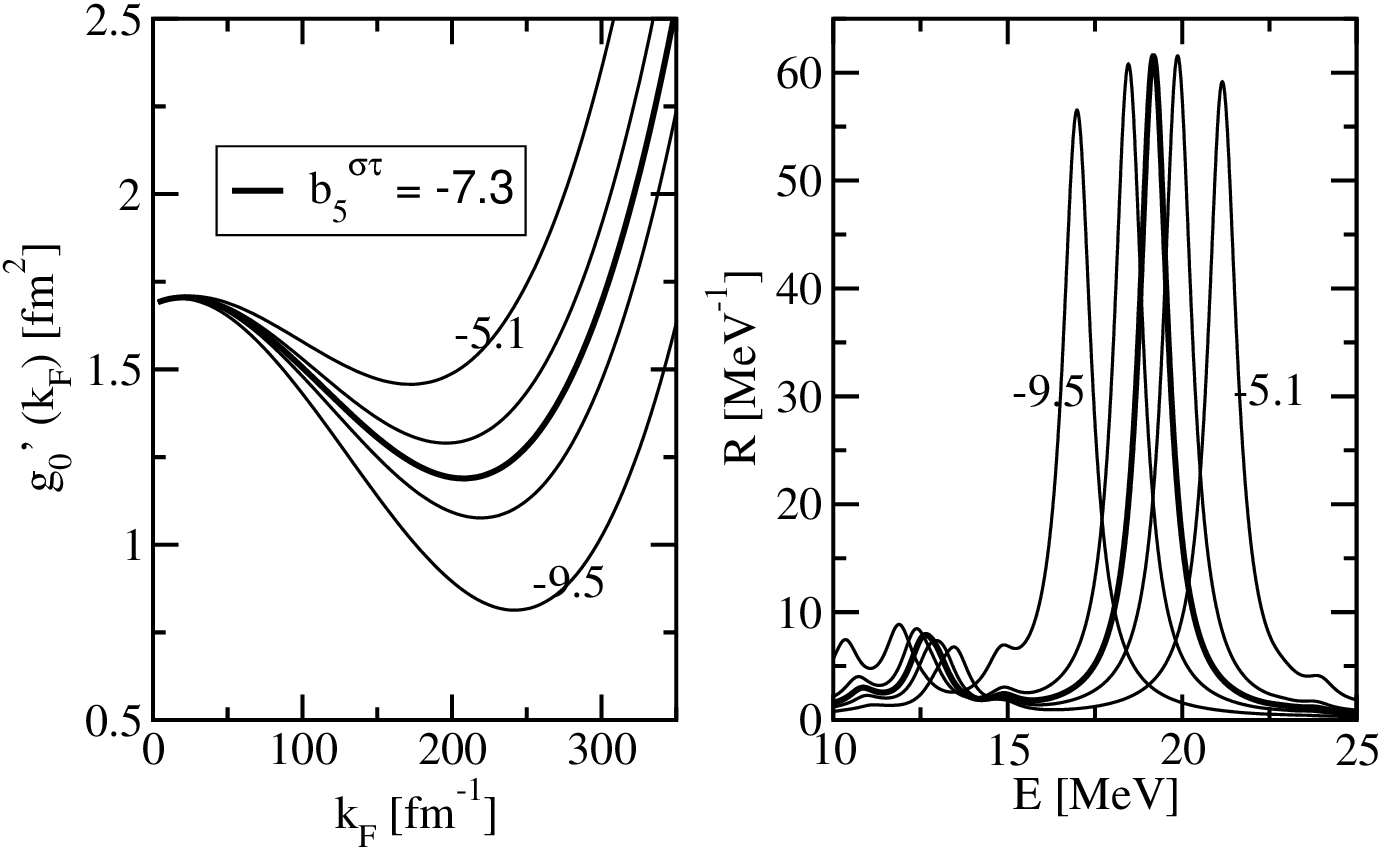}
\caption{\label{figA}
Left panel: the Landau-Migdal parameter $g_0'$ as 
function of the Fermi momentum $k_f$ for several values of 
$b_5^{\sigma \tau}$ in the interval $[-5.1,-9.5]$. 
In the panel on the right the corresponding Gamow-Teller 
strength functions in $^{208}$Pb are shown. The experimental
position of the GTR is reproduced with the choice: 
$b_5^{\sigma \tau} = -7.30$.}
\end{figure}

The density-dependent Landau-Migdal parameter $g_0'(k_f)$ of 
Ref.~\cite{Kaiser:2006wt} can be expanded in powers of the Fermi 
momentum $k_f$ to the same order as $G_{TS}^{(\pi)}(k_f)$ and 
$G_{TV}^{(\pi)}(k_f)$ \cite{Fi.06}:
\begin{equation}
\label{gprime}
g_0'(k_f) = c_1 + c_2 \frac{k_f}{\Lambda} + c_3 \left(
\frac{k_f}{\Lambda} \right)^2 + c_4 \left( \frac{k_f}{\Lambda}
\right)^3 \;,
\end{equation}
with the coefficients $c_i$ listed in Tab.~\ref{tab1}, and
$\Lambda = 2 \pi f_\pi \simeq 0.58$ GeV chosen as a reference scale. 
We note that, in addition to the terms which are completely 
determined by the in-medium ChPT calculation of pion-exchange 
diagrams, the constants $c_1$, $c_3$ and $c_4$ involve three
parameters as displayed in Tab.~\ref{tab1}:  
$b_3$, $b_5^{\sigma \tau}$ and $b_6^{\sigma \tau}$, respectively,
which encode the additional short-range dynamics of the 
$NN$ and $3N$ interactions. 
In Ref.~\cite{nucmat3} the subtraction constant 
$b_3=-3.05$ was adjusted to empirical nuclear matter properties at 
saturation point. Taking into account ground-state 
properties of finite nuclei in addition, this constant 
has been fine-tuned to $b_3=-2.93$ in Ref.~\cite{Fi.06}. 
This latter value will 
consistently be used both in the Dirac Hamiltonian which 
determines the nuclear ground state, and in the isovector 
residual RPA interaction Eq.~(\ref{Vph}). The remaining two 
short-distance constants $b_5^{\sigma \tau}$ and $b_6^{\sigma \tau}$ 
cannot be constrained by ground-state properties of spin-saturated
nuclear matter. In Ref.~\cite{Kaiser:2006wt} the value 
of $b_5^{\sigma \tau}$ has 
been adjusted so that the resulting $g_0'$ is consistent 
with empirical values at saturation density. Here we retain 
the value $b_6^{\sigma\tau}= -8.45$ from Ref.~\cite{Kaiser:2006wt}, 
and adjust $b_5^{\sigma \tau}$ in such a way that the PN-RPA 
calculation reproduces the excitation energy of the 
Gamow-Teller giant resonance (GTR) in $^{208}$Pb. In Fig.~\ref{figA} 
we plot the Landau-Migdal parameter  $g_0'$ as a
function of the Fermi momentum $k_f$ for several values of 
$b_5^{\sigma \tau}$ in the interval $[-5.1,-9.5]$. In 
the panel on the right the corresponding Gamow-Teller 
strength functions in $^{208}$Pb. The experimental position 
of the GTR in $^{208}$Pb: 19.2 MeV \cite{Aki.95,Hor.80,Kra.01}, 
is reproduced with $b_5^{\sigma \tau} = -7.30$. This
value is remarkably close to the estimate of Ref.~\cite{Kaiser:2006wt}.

\begin{table}
\begin{center}
\caption{\label{tab1}
Coefficients in the expansion of the spin-isospin
Landau-Migdal parameter $g_0'(k_f)$ Eq.~(\ref{gprime}). 
The best-fit constants used in the present 
calculation (middle column), are shown in comparison
with the estimates of Refs.~\cite{nucmat3,Kaiser:2006wt} (right column).
}
\vspace{0.5cm}
\begin{tabular}{c|c}
\hline
coefficients & constants \\
\hline
\begin{tabular}{c|c}

$c_1$ (fm$^2$) & ~ $-1.80~{\rm fm}^2 - \frac{\pi^2}{\Lambda^2} 
b_3$ \\
$c_2$ (fm$^2$) & ~ $1.17$ fm$^2$ \\
$c_3$ (fm$^2$) & ~ $-9.84~{\rm fm}^2 + \frac{\pi^2}{\Lambda^2} 
b_5^{\sigma \tau}$ \\
$c_4$ (fm$^2$) & ~ $21.13~{\rm fm}^2  - \frac{\pi^2}{\Lambda^2}
b_6^{\sigma \tau}$\\
\end{tabular}
&
\begin{tabular}{c|c}
$b_3               = -2.93$ & ~ $-3.05$ \cite{nucmat3}\\
 ~ & ~ \\
$b_5^{\sigma \tau} = -7.30$ & ~ $-8.03$ \cite{Kaiser:2006wt}\\
$b_6^{\sigma \tau} = -8.45$  & ~ $-8.45$  \cite{Kaiser:2006wt}\\
\end{tabular}\\
\hline
\end{tabular}
\end{center}
\end{table}

\begin{figure}
\includegraphics[scale=0.5,angle=0]{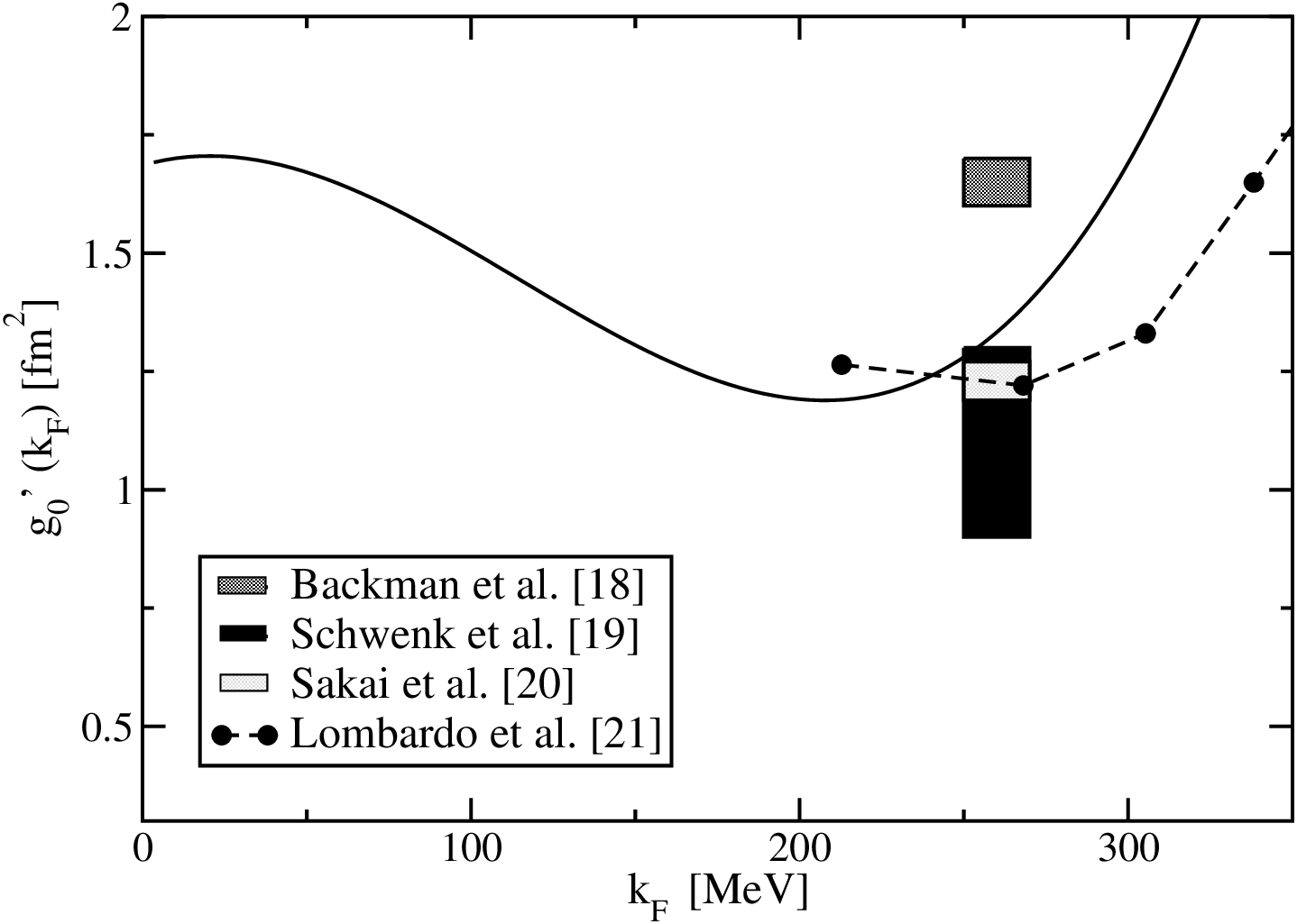}
\caption{\label{figB}
The dimensionfull Landau-Migdal parameter $g_0'$ 
as function of the Fermi momentum $k_f$ (solid curve). 
The values around saturation point are compared with 
those reported in Refs. \cite{backman} (dark gray rectangle),
\cite{schwenk} (black rectangle), and \cite{Sakai} 
(light gray rectangle). The black dots are the 
values obtained with the Brueckner-Hartree-Fock 
calculation of Ref.~\cite{Zuo}. The dimensionless 
$G_0'$ values from Refs.~\cite{backman,schwenk,Sakai,Zuo} 
have been divided by the density of states at the 
Fermi surface $N_0 = 2 \pi^{-2} k_{f0} M^* \approx 1~{\rm fm}^{-2}$.
}
\end{figure}

With the constants $b_3$, $b_5^{\sigma \tau}$ and $b_6^{\sigma \tau}$ 
adjusted as described above, we plot the resulting $g_0'$ as a
function of the Fermi momentum in Fig.~\ref{figB}. 
The values around saturation point ($k_f \approx 260$ MeV) 
are compared with: (a) a Brueckner calculation   
with a phenomenological $NN$-interaction \cite{backman} (dark gray rectangle);
(b) a recent estimate based on the universal low-momentum potential
$V_{low-k}$ \cite{schwenk} (black rectangle); (c) an empirical 
estimate based on the quenching of the Gamow-Teller strength \cite{Sakai} 
(light gray rectangle). We notice good agreement between the 
predicted values of $g_0'$ at saturation density, even though they 
have been obtained with very different models or estimated empirically. 
The functional dependence for $k_f > 200$ MeV is in qualitative 
agreement with the Brueckner-Hartree-Fock 
calculation of Ref.~\cite{Zuo} (black dots), 
which has employed the Argonne $V_{18}$ two-body potential 
\cite{Argonne18}, and a three-body force from Ref.~\cite{3body}. 

\section{Applications: spin-isospin collective excitations}

Collective spin and isospin excitations in nuclei
have been the subject of numerous experimental and theoretical
studies (for extensive reviews see Refs.~\cite{Ost.92,HW.01}). 
Nucleons with spin up and spin down can oscillate either in phase 
(spin scalar S=0 mode) or out of phase (spin vector S=1 mode).
The spin vector, or spin-flip excitations can be of isoscalar
(S=1, T=0) or isovector (S=1, T=1) nature. These modes provide 
information on the spin and spin-isospin
channels of effective nuclear interactions. 

\subsection{Isobaric analog resonances}

\begin{figure}
\includegraphics[scale=0.58,angle=0]{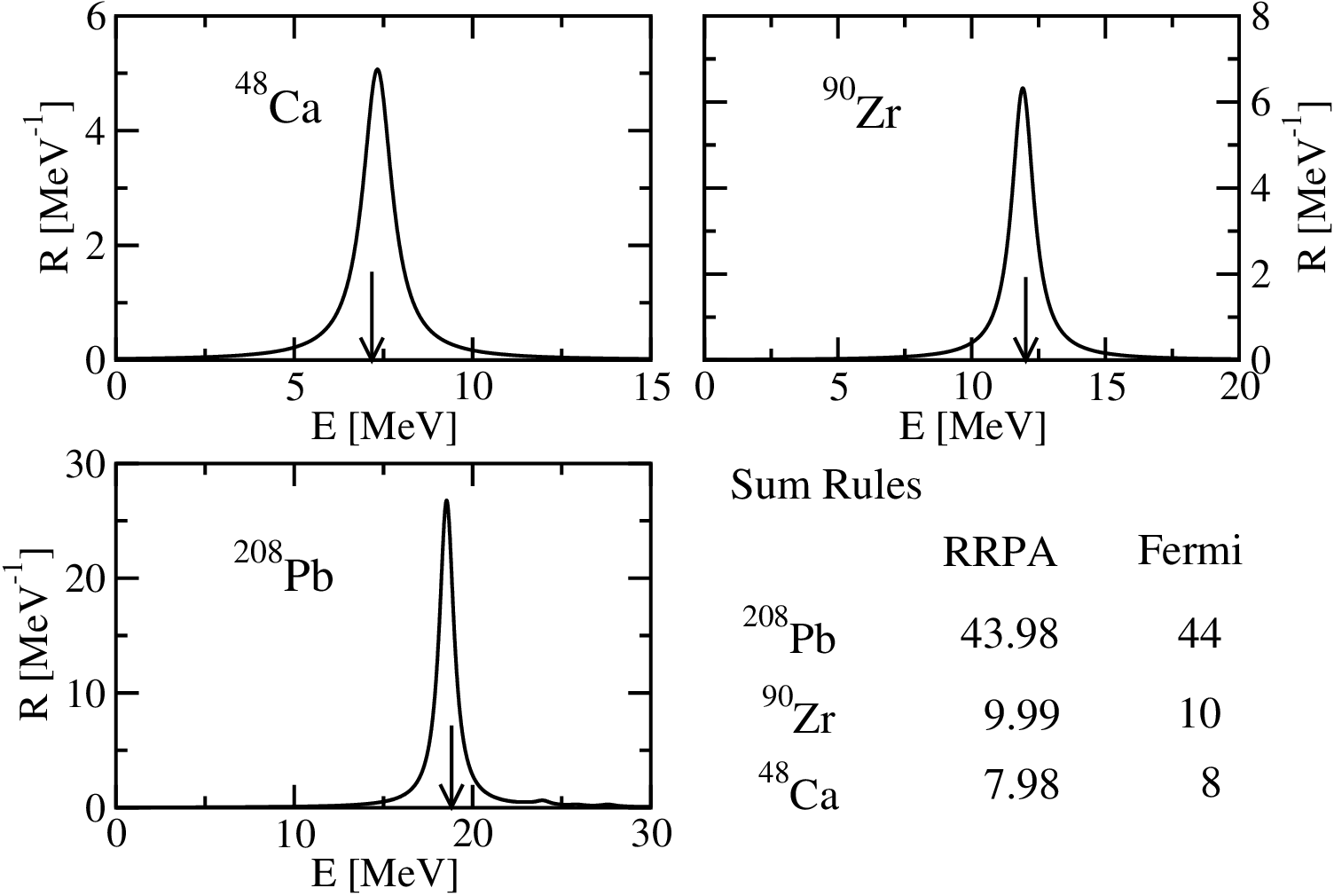}
\caption{\label{figC} 
PN-RRPA $J^\pi=0^+$ strength distributions. 
The excitations of the isobaric analog resonances 
are compared with  experimental data (arrows) 
for $^{48}$Ca~\protect\cite{And.85}, $^{90}$Zr~\protect\cite{Wakexp.97}, 
and $^{208}$Pb~\protect\cite{Aki.95,Bai.80}.
The integrated strengths are compared to the model-independent
Fermi sum rule Eq.~(\ref{iasrule}). 
}
\end{figure}

As a first application of this new PN-QRPA model, we calculate the 
strength functions for the simplest charge-exchange mode: the 
isobaric analog resonance (IAR) $J^{\pi}=0^{+}$. 
The one-body Fermi transition operator reads:
\begin{equation}
T_{\beta^{\mp}}^{F}=\sum_{i=1}^{A}\tau_{\mp}(i)\; .
\label{iaroperator}
\end{equation}
For $N > Z$ nuclei $T_{\beta^{-}}^{F}$ simply changes a neutron 
into a proton without spin-flip or change in orbital angular momentum. 
In Fig.~\ref{figC} we plot the calculated PN-QRPA response to the operator
Eq.~(\ref{iaroperator}) for $^{48}$Ca, $^{90}$Zr and $^{208}$Pb. 
The strength distributions are dominated by a single
IAR peak which corresponds to a coherent superposition of 
proton-particle -- neutron-hole excitations. 
The calculated IAR excitation energies 
(evaluated with respect to the calculated 
ground state of the parent nucleus) are 
compared with the corresponding experimental values from 
$(p,n)$ charge-exchange scattering data for
$^{48}$Ca ($7.2$ MeV) \cite{And.85}, $^{90}$Zr ($12.0$ MeV) \cite{Wakexp.97}, 
and $^{208}$Pb ($18.8$ MeV)\cite{Bai.80}. 
The agreement between the PN-QRPA results and experimental 
data is very good. We have also verified that the calculated strength 
distributions exhaust the Fermi sum rule 
\begin{equation}
S_{\beta^{-}}^{F}  = 
\sum_f |\langle \psi_f| T_{\beta^{-}}^{F} | \psi_i \rangle|^2 = 2
\langle \psi_i | T_3 | \psi_i \rangle = N-Z \; ,
\label{iasrule}
\end{equation}
to better than 99.75\% (see Fig.~\ref{figC}).

\begin{figure}
\includegraphics[scale=0.58,angle=0]{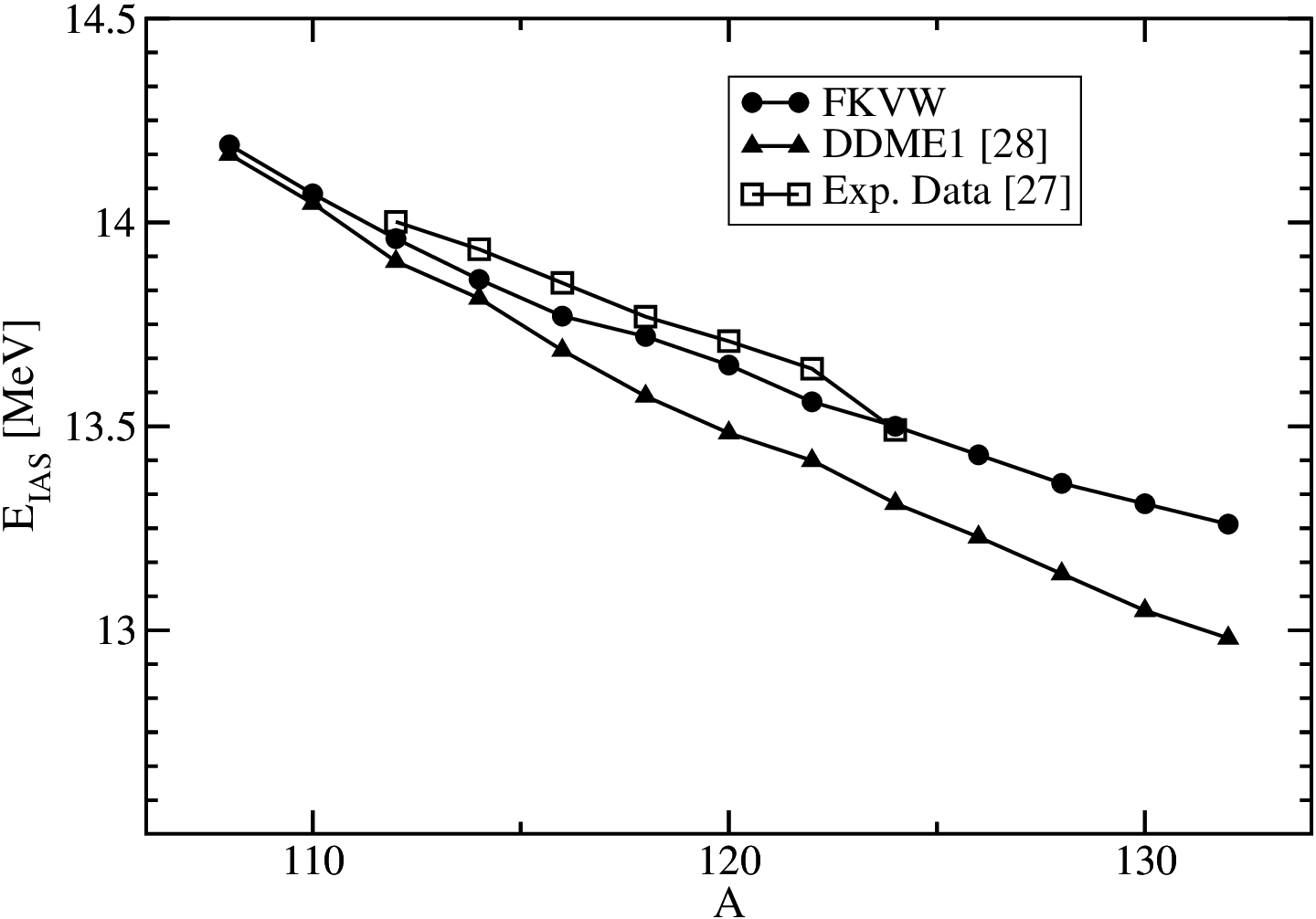}
\caption{\label{figD} RHB plus PN-QRPA results for the 
isobaric analog resonances of the sequence of even-even Sn target 
nuclei, calculated with the FKVW + Gogny effective interaction and 
compared with experimental data~\cite{Pham.95}, and with the 
calculation of Ref.~\cite{Paar:2004re}. 
}
\end{figure}

As a second example we consider open-shell nuclei and plot, in
Fig.~\ref{figD}, the calculated IAR excitation energies for the
sequence of even-even Sn target nuclei with $A=108 - 132$. The 
result of our self-consistent RHB plus PN-QRPA 
calculation are shown in comparison with experimental data 
obtained in a systematic study of the ($^3$He,t) charge-exchange 
reaction over the entire range of stable Sn isotopes \cite{Pham.95}, 
and with the relativistic PN-QRPA analysis of Ref.~\cite{Paar:2004re},
in which the phenomenological density-dependent 
meson-exchange interaction DD-ME1 \cite{DDME1} was used both 
in the RHB calculation of the ground states and in the particle-hole
($ph$) channel of the isovector residual interaction. Both models 
reproduce the empirical mass dependence of the IAR, and  
the calculated excitation energies are in very good agreement 
with available data. For DD-ME1 the largest difference between 
the theoretical and experimental IAR excitation energies is 
$\approx 200$ keV, and an even better agreement with data is 
obtained in the present calculation. This is because the 
FKVW effective interaction reproduces the ground 
states of Sn isotopes (e.g. binding energies) better 
than DD-ME1 \cite{Fi.06}. We note that, because the IAR is a 
non spin-flip mode, the one-pion exchange Eq.~(\ref{one-pi}) 
and the Landau-Migdal term Eq.~(\ref{LM}) do not contribute 
to the matrix elements of the residual interaction Eq.~(\ref{Vph}). 
The only contribution comes from the isovector channel of the 
FKVW interaction (i.e. the Lagrangian density Eq.(\ref{FKVW-IV})). 
The density-dependent coupling strengths $G_{TS}^{(\pi)}$ and
$G_{TV}^{(\pi)}$ have been adjusted to the (asymmetric) nuclear 
matter equation of state and ground-state properties of finite nuclei.

We would also like to emphasize a result that was already discussed 
in Ref.~\cite{Paar:2004re}, namely that sharp, non-fragmented IAR 
peaks in open-shell nuclei are only obtained when the $T=1$ pairing 
interaction is consistently included both in the RHB calculation of 
the nuclear ground-state and in the proton-neutron residual interaction.
This is simply a consequence of the fact that the pairing interaction
is isospin invariant, and therefore commutes with the Fermi
operator Eq.~(\ref{iaroperator}).

We note that the Nolen-Schiffer anomaly~\cite{Nolenschiffer} is absent
in the present calculation because the attractive Coulomb exchange term
is not included in the energy density functional that we use 
in the calculation of ground state properties~\cite{Brown:2000di}.
Further refinements should also consider 
charge symmetry breaking contributions,
e.g. from iterated and irreducible 
pion-photon exchange processes~\cite{Kaiser:2003ty}.

\subsection{Gamow-Teller excitations}

\begin{figure}
\includegraphics[scale=0.58,angle=0]{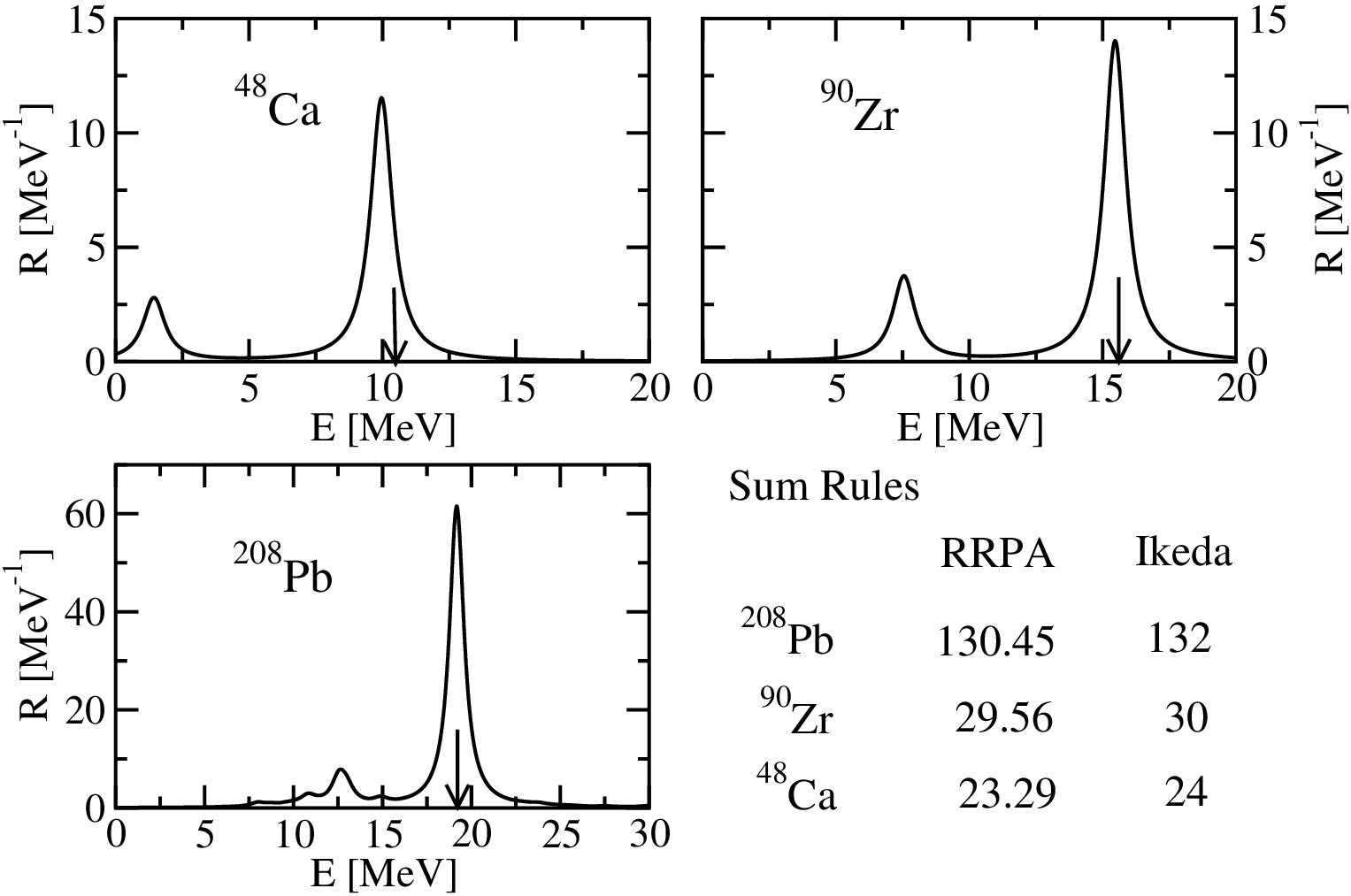}
\caption{\label{figE}
Gamow-Teller 
strength distributions for 
$^{48}$Ca, $^{90}$Zr and $^{208}$Pb. PN-RQPA results are 
shown in comparison with experimental data (arrows) for the 
GTR excitation energies in $^{48}$Ca~\protect\cite{And.85}, 
$^{90}$Zr~\protect\cite{Wakexp.97,Bai.80}, 
and $^{208}$Pb~\protect\cite{Aki.95,Hor.80,Kra.01}.
}
\end{figure}

The calculated Gamow-Teller ($J^{\pi}=1^{+}$) strength distributions 
for $^{48}$Ca, $^{90}$Zr and $^{208}$Pb are shown in Fig.~\ref{figE}. 
The one-body Gamow-Teller operator reads:
\begin{equation}
T_{\beta^{\mp}}^{GT}=\sum_{i=1}^{A}\bm{\Sigma}(i)\tau_{\mp}(i) \; .
\label{gtroperator}
\end{equation}
The corresponding integrated strengths satisfy the Ikeda sum rule: 
\begin{equation}
S_{\beta^{-}}^{GT}-S_{\beta^{+}}^{GT} =
\sum_f | \langle \psi_f | T_{\beta^{-}}^{GT} | \psi_i \rangle |^2 -
\sum_f | \langle \psi_f | T_{\beta^{+}}^{GT} | \psi_i \rangle |^2 
= 3(N-Z) \; .
\label{gtsrule}%
\end{equation}
In addition to the high-energy GT 
resonance -- a collective superposition of direct spin-flip 
($j = l + \frac{1}{2}$ $\rightarrow$ $j = l - \frac{1}{2}$) 
transitions -- the response functions display a concentration of
strength in the low-energy tail. The transitions in the low-energy
region correspond to core-polarization 
($j = l \pm \frac{1}{2}$ $\rightarrow$ $j = l \pm \frac{1}{2}$), 
and back spin-flip 
($j = l - \frac{1}{2}$ $\rightarrow$ $j = l + \frac{1}{2}$)
neutron-hole -- proton-particle excitations. The calculated 
GTR are compared with the experimental excitation energies: 
$10.5$ MeV for $^{48}$Ca \cite{And.85}, $15.6$ MeV for 
$^{90}$Zr \cite{Wakexp.97,Bai.80}, and $19.2$ MeV for 
$^{208}$Pb~\cite{Aki.95,Hor.80,Kra.01}. 
Although one of the parameters of the Landau-Migdal 
interaction has been adjusted to reproduce
the GTR excitation energy in $^{208}$Pb, we find a very good
agreement with experiment also for $^{48}$Ca and $^{90}$Zr.
The integrated strengths satisfy the Ikeda sum rule with 
high accuracy. This is an important test of the 
internal consistency of our relativistic PN-RPA. We note that 
the Ikeda sum rule is exhausted
by the calculated GT strength only when the relativistic RPA/QRPA 
space includes both the $ph$ excitations formed from ground-state 
configurations of the fully or partially occupied states of
positive energy, and the empty negative-energy states from the 
Dirac sea \cite{Paar:2004re}. The contribution of these
configurations to the Ikeda sum rule is of the order of $8-10$\%. 

\begin{figure}
\includegraphics[scale=0.58,angle=0]{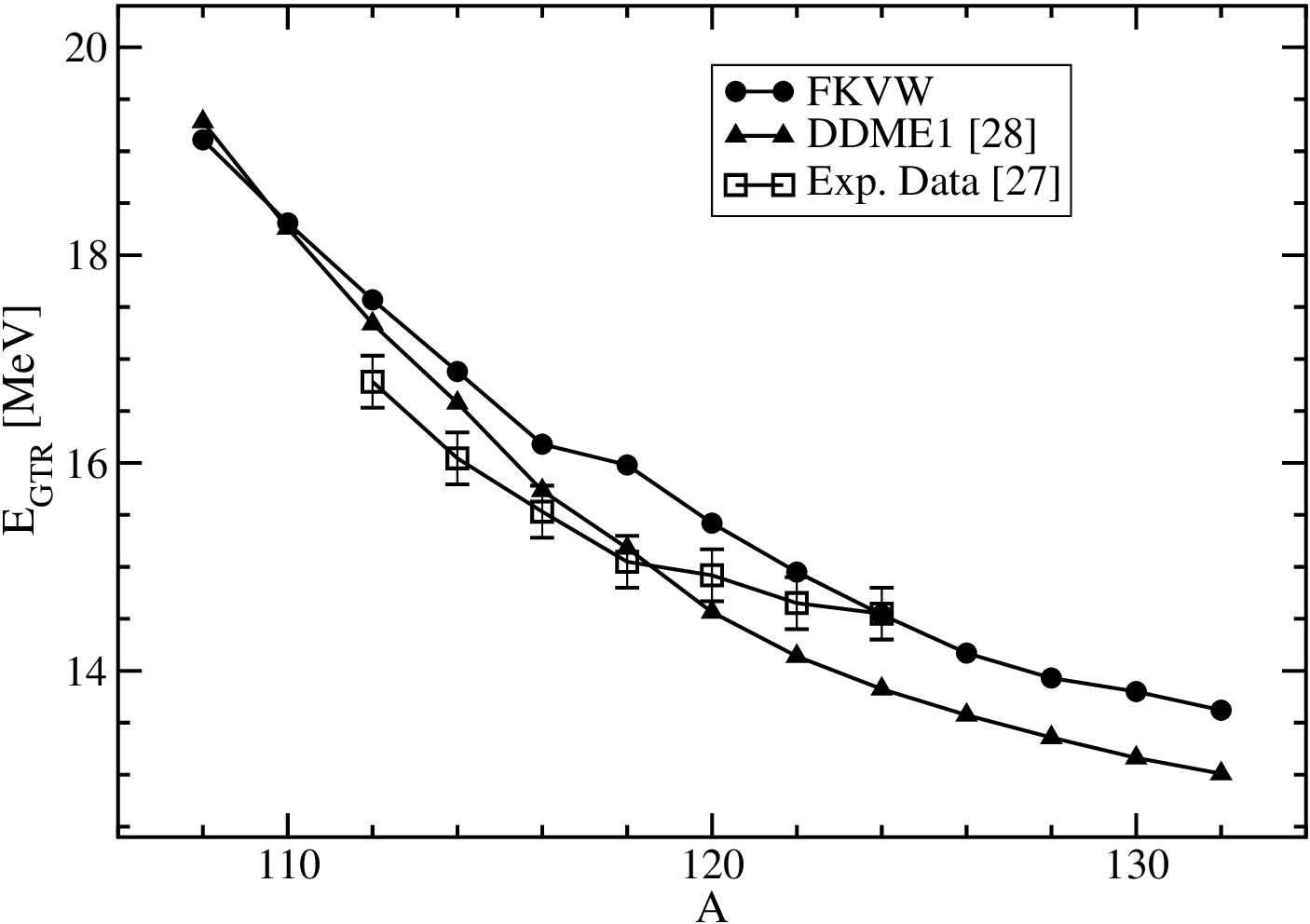}
\caption{\label{figF}
Excitation energies of Gamow-Teller resonances for the 
sequence of even-even $^{108 - 132}$Sn target nuclei. 
The present relativistic PN-QRPA calculation is compared 
with the results of Ref.~\protect\cite{Paar:2004re}, and
with experimental data from Ref.~\protect\cite{Pham.95}.
}
\end{figure}

Finally, for the sequence of even-even Sn target nuclei, 
we compare in Fig.~\ref{figF} the PN-QRPA predictions 
for the GTR excitation energies with experimental data 
from Sn$(^3$He,t$)$Sb charge-exchange reactions \cite{Pham.95}, 
and with the results obtained with the DD-ME1 meson-exchange 
effective interaction in Ref.~\cite{Paar:2004re}. The same $T=1$ 
Eq.~(\ref{GognyT1}) and $T=0$ Eq.~(\ref{GaussT0}) 
pairing interactions have been used in both PN-QRPA calculations. 
Both models reproduce the isotopic trend of GTR 
excitation energies.
For the individual nuclei the level of agreement with data 
varies. Below $A=120$ the GTR calculated with DD-ME1 plus the 
residual interaction of Ref.~\cite{Paar:2004re} are closer to 
the experimental energies, whereas for $A>120$
the present PN-QRPA calculation using the FKVW parameterization 
plus the residual interaction Eq.~(\ref{Vph}) predicts GTR in better 
agreement with data.

\section{Conclusions}

In summary, we have extended our approach to nuclear density functional 
theory, based on chiral dynamics and the symmetry breaking pattern 
of low-energy QCD, to the description of charge-exchange 
excitations in finite nuclei.  Starting from the relativistic 
nuclear density functional (FKVW) introduced in Ref.~\cite{Fi.06},  
the proton-neutron (quasiparticle) random phase approximation has 
been formulated and applied in a study of the role of chiral 
pion-nucleon dynamics in excitation modes which include spin and 
isospin degrees of freedom. In addition to the isovector channel 
of the FKVW effective interaction, the RPA residual interaction 
includes the direct one-pion exchange and the zero-range 
Landau-Migdal term. 

The density dependence of the Landau-Migdal 
parameter ${g_0'(k_f)}$ which characterizes the spin-isospin 
channel of the isotropic part of the quasi-nucleon interaction
at the Fermi surface, has been calculated in Ref.~\cite{Kaiser:2006wt} 
using in-medium ChPT. 
In the present work one of the short-distance constants 
has been fine-tuned in such a way that the relativistic 
PN-QRPA calculation reproduces the excitation energy of the 
Gamow-Teller resonance (GTR) in $^{208}$Pb. 
Both approaches lead to very similar values of ${g_0'(k_f)}$.
In addition to the 
residual interaction in the $ph$ channel, the model includes 
both the $T=1$ and $T=0$ pairing channels. PN-QRPA 
calculations have been performed for the $J^{\pi}=0^{+}$ and 
$J^{\pi}=1^{+}$ charge exchange modes in $^{48}$Ca, 
$^{90}$Zr and $^{208}$Pb, and for a sequence of even-even 
$^{108 - 132}$Sn nuclei. Results for the excitation energies 
of IAR and GTR, and especially for the isotopic trend in the chain 
Sn nuclei, are in very good agreement with available data and 
with calculations performed using modern phenomenological 
relativistic density-dependent meson-exchange effective interactions.

This analysis has shown that a nuclear energy density functional 
based on chiral effective field theory provides a consistent 
microscopic framework not only for ground-state properties, 
but also for complex excitations of the nuclear many-body system. 
The specific density dependence of spin-isospin dependent 
particle-hole interactions, as derived from in-medium chiral dynamics,
gives a successful description of both isobaric analogue and
Gamow-Teller modes.

\bigskip 
\leftline{\bf ACKNOWLEDGMENTS}
We thank Tamara Nik{\v{s}}i{\'{c}} and Nils Paar for helping us 
with the development of the relativistic QRPA codes, and Peter 
Ring for useful discussions. This work has been supported in part by
BMBF, DFG, GSI, MURST, INFN, and MZOS.


\newpage


\newpage

\newpage

\newpage

\newpage

\newpage

\end{document}